\documentstyle[prl,aps,psfig]{revtex}
\newcommand{\be}{\begin{equation}}
\newcommand{\ee}{\end{equation}}
\newcommand{\bea}{\begin{eqnarray}}
\newcommand{\eea}{\end{eqnarray}}
\newcommand{\p}{\prime}
\newcommand{\li}{\int_{0}^{\infty}}
\newcommand{\ri}{\int_{0}^{1}}

\def\q{$ {\bf p}$}

\begin{document}
\title{Conserved mass models and particle systems in one dimension}
\author{R. Rajesh and Satya N. Majumdar}
\address{Department of Theoretical Physics, Tata Institute of Fundamental
Research,
Homi Bhabha Road, Bombay 400005, India}
\date{\today}
\maketitle
\widetext
\begin{abstract}
In this paper we study analytically a simple one dimensional model of mass
transport. We introduce a parameter $p$ that interpolates between
continuous time dynamics ($p\to 0$ limit) and discrete parallel update
dynamics ($p=1$). For each $p$, we study the model with (i) both
continuous and discrete masses and (ii) both symmetric and asymmetric
transport of masses. In the asymmetric continuous mass model, the two
limits $p=1$ and $p\to 0$ reduce respectively to the $q$-model of force
fluctuations in bead packs [S.N. Coppersmith et. al., Phys. Rev.  E. {\bf
53}, 4673 (1996)] and the recently studied asymmetric random average
process [J. Krug and J. Garcia, cond-mat/9909034]. We calculate the steady
state mass distribution function $P(m)$ assuming product measure and show
that it has an algebraic tail for small $m$, $P(m)\sim m^{-\beta}$ where
the exponent $\beta$ depends continuously on $p$.  For the asymmetric case
we find $\beta(p)=(1-p)/(2-p)$ for $0\leq p <1$ and $\beta(1)=-1$ and for
the symmetric case, $\beta(p)=(2-p)^2/(8-5p+p^2)$ for all $0\leq p\leq 1$. 
We discuss the conditions under which the product measure ansatz is exact. 
We also calculate exactly the steady state mass-mass correlation function
and show that while it decouples in the asymmetric model, in the symmetric
case it has a nontrivial spatial oscillation with an amplitude decaying
exponentially with distance.
\end{abstract}


\section{Introduction} 
There is a wide variety of physical systems in nature where the basic
microscopic dynamical processes involved are aggregation, fragmentation,
adsorption, desorption and transport of mass.  These processes are
abundant and occur in systems such as colloidal suspensions\cite{White},
polymer gels\cite{Ziff,Krapiv}, river networks\cite{river}, aerosols and
clouds\cite{Fried} and surface growth phenomena involving island
formation\cite{Lewis}.  These systems can have different types of
non-equilibrium stationary states and phase transitions between them as
the rates of the underlying microscopic processes are varied. While for
systems in thermal equilibrium the stationary state is characterized by
the Gibbs measure, there is no such general recipe for non-equilibrium
systems.  In order to gain more insights on the nature of these steady
states and possible phase transitions between them, several simple lattice
models involving mass transport have been proposed and studied
recently\cite{Tak,MKB,MKB1}. By virtue of their simplicity, these lattice
models are often amenable to exact analysis and yet contain rich and
nontrivial physics.

These models constitute simple examples of interacting many particle
systems out of equilibrium; in particular the dynamics of these systems do
not obey detailed balance.  The steady states of interacting many body
systems are in general difficult to characterize and only a few exact
results are available.  These include simple exclusion processes with open
and closed boundary conditions \cite{derrida}, abelian sandpile models of
self organized criticality \cite{dhar}, traffic models \cite{cates} and
mass aggregation model of Takayasu \cite{Tak}.  Moreover the steady states
in some cases are non universal and depend on the detailed nature of the
dynamics used for updating. For example the steady state in the asymmetric
simple exclusion process depends on whether the update rules are parallel
or random sequential \cite{evans}. It is therefore desirable to study more
of such simple models in a systematic way in order to get insight into the
nature of the non-equilibrium steady states. In this paper we study a
simple lattice model of mass transport analytically which sheds some light
on these general issues pertaining to interacting many body systems. 

Besides, the dynamics in seemingly unrelated systems can often be mapped
onto simple one dimensional mass transport models evolving with time
according to some prescribed rules. These systems include river
networks\cite{Tak,AM}, force fluctuations in granular systems such as bead
packs\cite{SC}, traffic flows\cite{krug}, voting systems\cite{Melzak,FF},
wealth distributions\cite{IKR}, generalized Hammersley
process\cite{aldous} and inelastic collisions in granular
gases\cite{Haff}.

In this paper we study analytically a model of mass transport in a one
dimensional lattice. Each lattice site contains a nonnegative mass
variable and the dynamics consists of transporting a finite amount of mass
from each site to its neighbours. The amount to be transported is randomly
chosen from a given distribution. We introduce a parameter $p$ that
interpolates between continuous time dynamics ($p\to 0$ limit) and
discrete parallel update dynamics ($p=1$). For each $p$, we study the
model with (i) both continuous and discrete masses and (ii)  both
symmetric and asymmetric transport of masses. 

The paper is organized as follows. In section II, we define the mass model
precisely and discuss its mapping to other models of non equilibrium
statistical mechanics. We also summarize our main results.  In section
III, we discuss the asymmetric continuous mass model and solve for the
mass distribution function in the steady state assuming that product
measure holds.  We then discuss under what conditions the product measure
is exact.  The two point mass correlation function is also computed
exactly.  In section IV, we study the symmetric version of the continuous
mass model. We show that product measure becomes exact in a particular
limit.  We also compute the stationary two point mass correlation function
exactly for the symmetric model.  In section V, we study the discrete mass
version of the model.  Finally we conclude with a summary and outlook in
section VI.  Appendix A contains a proof that product measure fails for
the asymmetric model for any $p<1$. In appendix B we prove that product
measure is exact for the symmetric model in the $p\to 0$ limit.

\section{The model}

Our mass model is defined on a lattice. For simplicity, we define it here
on a one dimensional lattice with periodic boundary conditions. The
generalization to higher dimensions is straightforward. Each lattice site
contains a nonnegative mass variable. We consider two versions of the
model: (i) when the mass at each site is a continuous variable and (ii)
when the mass at each site is discrete. 

First consider the continuous mass model (see figure 1). We start with a random
configuration of masses, $m(i)$ at each site $i$. The dynamics is defined
as follows. Each discrete time step consists of two moves: (1) a fraction
$\eta_i$ of each mass $m_i$ is chosen and then (2) with probability $p$,
this fraction is added to a neighbouring site and with probability $(1-p)$
it remains at the original site. In the asymmetric model, the fraction is
always added to the right neighbour. In the symmetric case, the fraction
goes either to the left or the right neighbour with equal probability.  We
note that total mass is conserved by the dynamics. Thus the model has two
parameters, the probability $p$ and the average mass per site $\rho$. 

The fractions $\eta_i$ are independent and identically distributed random
variables drawn from a probability distribution on $[0,1]$. In this paper,
we will mostly consider a uniform distribution of $\eta_i$ on $[0,1]$
though some of our results can be generalized to a class of other
distributions. 

We note that for $p=1$, all the chosen fractions of masses are definitely
transported to their neighbours. This corresponds to fully parallel update
dynamics. In this case, the asymmetric version of the model reduces
exactly to the $q$-model introduced by Coppersmith et. al.\cite{SC} to
study force fluctuations in random bead packs.  In this case, the mass at
each site evolves as 
\bea 
{m_i}(t+1)=\eta_{i-1}m_{i-1}(t)+(1-\eta_i)m_i(t).  
\label{one} 
\eea 
In the context of bead packs, the indices $i$ and $t$ index the site $i$ 
at a depth $t$ in a two dimensional packing. Then for large $t$, $m_i$ 
represents the force supported by a bead at $(i,t)$ scaled by the mean 
weight and $\eta_im_i$ is the random component of the weight (scaled by 
the mean weight) transmitted from a bead at depth $t$ to a neighbouring 
one at depth $(t+1)$ that touches it. The same equation was
also studied in Ref.\cite{MKB1} in the context of a lattice gas model. The
stationary mass distribution $P(m)$ of the $q$-model was solved
exactly\cite{SC} and remarkably the mean field theory turned out to be
exact in the thermodynamic limit for the case of uniform distribution of
the fractions $\eta_i$'s. This means that the steady state joint
distribution of masses at different sites factorises,
$P(m_1,m_2,m_3\ldots)=\prod_i P(m_i)$. In other words, the product measure
is exact in this case and $P(m)$ was shown\cite{SC,MKB1} to have a simple
distribution, 
\bea 
P(m)={4m\over {\rho^2}}e^{-2m/\rho}.  
\label{two} 
\eea
   
In the opposite limit $p\to 0$, the probability that two or more sites will
be simultaneously updated in a single move is $O(p^2)$ and hence negligible. 
With the choice of $p=\Delta t$, this limit thus corresponds to the random 
sequential continuous time dynamics. This case has been studied recently by 
Krug and Garcia\cite{krug} and {\it assuming} that product measure holds 
they derived the steady state mass distribution $P(m)$ for uniform
distribution of $\eta_i$'s,
\bea
P(m)=\frac{1}{\sqrt{2 \pi \rho m}} \exp(-m/{2 \rho}).
\label{three}
\eea

The difference in the small $m$ behaviour of the mean field $P(m)$ in the
parallel and random sequential case was correctly noted by Krug and
Garcia\cite{krug}. We have studied both the symmetric and asymmetric
versions of the model for arbitrary $p$. Our main results are summarized
as follows: 

(1) In the asymmetric case, we show that within the mean field
approximation $P(m)\sim m^{-\beta}$ for small $m$, while it decays
exponentially for large $m$. The exponent $\beta(p)$ depends continuously
on $p$ with a discontinuity at $p=1$, $\beta(p)=(1-p)/(2-p)$ for $p<1$ and
$\beta(1)=-1$. In the symmetric case, the mean field $P(m)$ has a similar
behaviour except the exponent $\beta(p)=(2-p)^2/(8-5p+p^2)$ for all $p$ in
$[0,1]$. Note that in the symmetric case, there is no discontinuity at
$p=1$.
 
(2) In the asymmetric case, we prove rigorously that the product measure
is exact only for $p=1$. For any $p<1$ (including the random sequential
$p=0$ case), we show that the product measure ansatz,
$P(m_1,m_2,m_3,\ldots)=\prod_i P(m_i)$ breaks down. But remarkably the
mean field $P(m)$ is almost indistinguishable from the $P(m)$ obtained
from numerical simulation in one dimension. We note that the breakdown of
product measure property does not necessarily mean that the correct single
point distribution $P(m)$ is still not given by the mean field $P(m)$;  in
fact numerical results strongly suggest that the mean field $P(m)$ is
exact even though product measure fails. In the symmetric case on the
other hand, product measure is exact only for $p\to 0$ but fails for any
$p>0$. Besides, as opposed to the asymmetric case, the mean field $P(m)$
is considerably different from the distribution obtained numerically. This
is due to strong correlations between masses in the symmetric case as
mentioned below. 

(3) The two-point mass correlation function $C(r)=\langle m(0)m(r)\rangle$
between two sites at distance $r$ can be computed exactly (without
recourse to the assumption of product measure) for arbitrary $p$ in both
asymmetric and symmetric models. We find that in the asymmetric case, for
all values of $p$, the connected part of the correlation function
vanishes, $C(r)-\rho^2=0$ for $r>0$. In the limit $p\to 0$, this fact was
noted by Krug and Garcia\cite{krug}. This however does not imply the
validity of product measure is exact which would require factorization of
all higher order correlations as well. For the symmetric case, the
correlation function factorises only for $p=0$. However for $p>0$, the
function $C(r)-\rho^2$ has a nontrivial spatial dependence.  It oscillates
with distance $r$ and the amplitude of the oscillation decays
exponentially with $r$. 

We have also studied a discrete mass version of the above model.  In this
case the mass $m_i$ at any site $i$ can take only discrete non-negative
integer values. Instead of a random fraction breaking off a mass as in the
continuous case, the mass to be taken out of a site is a random variable
that takes only discrete values $0$,$1$,$2$,$3\ldots$ , $m_i$ equally
likely, i.e., any of these values is chosen with the same probability
$1/(m_i+1)$. Then as before, with probability $p$, the chosen mass is
actually transported to a neighbouring site and with probability $(1-p)$
it stays at its original site. We derive the explicit expressions for the
mass distribution for the discrete case also. 
   
This model can be mapped onto a model of hard core particles moving with
long range jumps in one dimension\cite{MKB1,krug}. First consider the
continuous mass model.  Each site of the lattice corresponds to a particle
(point) on the real line and the mass $m_i$ represents the continuous gap
between $i$-th and $(i+1)$-th particle. The transport of random fraction
of $m_i$ from the $i$-th site to $(i+1)$-th site corresponds to the
$(i+1)$-th particle jumping to the left by a random fraction of the
available gap between it and its left neighbour (see figure 2).  The
discrete mass problem similarly corresponds to particles moving on a one
dimensional lattice (as opposed to the real line in the continuous case)
with hard core repulsion. At each time step a particle moves to a site
randomly chosen from the set of empty sites in front of it. This is a
generalization of the simple exclusion process where a particle can jump
only to a nearest neighbour site provided it is unoccupied. In this
generalized case, while the hard core repulsion is respected, long range
jumps are allowed. 

The discrete mass problem can also be mapped onto an interface growth
problem via the usual mapping from a lattice gas model to a growing
interface. Starting from a reference height $h=0$, a particle at site $i$
corresponds to $h(i+1)=h(i)-1$ while a hole corresponds to
$h(i+1)=h(i)+1$. Under this mapping, our problem corresponds to the
following rules: Any stretch of the interface with slope equal to $1$ can
be split at any randomly chosen point in between into two sections of
slope $1$ connected by a bond of slope $-1$. 

\section{The asymmetric model with continuous mass }

In this section, we study the continuous mass model where in each time
step a fraction of the mass from any given site is transported with
probability $p$ to its right neighbour. The
Langevin equation for the mass update can be written as
\bea
m_i(t+1)=m_{i}(t)-\sigma_i\eta_i m_i(t)+\sigma_{i-1}\eta_{i-1}m_{i-1}(t),
\label{four}
\eea
where the fractions $\eta_i$'s are random numbers in $[0,1]$ chosen from
a uniform distribution and
the random variables $\sigma_i$'s  take values $1$ with probability $p$ or
$0$ with probability $q=1-p$. The distribution of both of these variables
are independent from site to site. Defining $r_i=\eta_i \sigma_i$, 
we get
\bea
m_i(t+1)=m_i(t)(1-r_i)+m_{i-1}(t) r_{i-1}, 
\label{five}
\eea
and it is not difficult to see that the effective distribution $f(r_i)$ of
the random variable $r_i$ on $[0,1]$ given by
\bea
f(r_i)= q\delta(r_i)+p .
\label{six}
\eea

The evolution equation of the single point mass distribution function
$P(m_i,t)$ (which is
independent of $i$ due to translational invariance) can be written
down exactly,
\bea
P(m_i,t+1)&=&
\li dm_{i-1} \ri dr_{i-1} f(r_{i-1}) \li dm^{\p}_{i} \ri dr_i f(r_i)
P(m_{i-1},m_{i}^{\p},t) \nonumber \\
&\times&\delta (m^{\p}_i (1-r_i)+m_{i-1} r_{i-1}-m_i).
\label{seven}
\eea
Here $P(m_{i-1},m_i,t)$ is the joint probability distribution of mass
$m_{i-1}$
at site $i-1$ and $m_i$ at site $i$. 
The time evolution of the single point
probability distribution involves the two point joint probability
distribution function. Similarly the n-point probability distribution will
involve the (n+1)-point joint probability distribution
and in general this hierarchy cannot be broken.

\subsection{Mean Field Theory}
We first compute the single point mass distribution
$P(m)$ from Eq. (\ref{seven}) by assuming that the joint 
distribution factorises in the steady state,
$P(m_{i-1},m_{i})=P(m_{i-1})P(m_i)$. This approximation clearly ignores
correlations between masses. Within this approximation, Eq. (\ref{seven})
involves only single point
distribution function $P(m)$. Taking the stationary limit, $t\to \infty$,
and using the explicit form of the distribution $f(r)$ from Eq.(\ref{six}), we
find that the Laplace
transform, $Q(s)=\int_0^{\infty} P(m)\exp(-ms)dm$
satisfies the equation,
\bea
Q(s)=\frac{p V (p V+q)}{1-p q V- q^2} 
\label{eight}
\eea
where $V(s)=\ri Q(s r)dr$. We note that $\frac{d}{ds} (s
V)=Q(s)$.
Eliminating $Q$, we get a first order differential equation for $V$ which
can be integrated to give,
\bea
\frac{1-V}{V^{2-p}}=\rho s/2,
\label{nine}
\eea
where the integration constant has been determined by using the
fact that $dQ/ds|_{s=0}=-\rho$ with $\rho$ being the average mass per
site. The above equation reduces to a quadratic, cubic and linear equation in
$V$ for $p=0$, $p=0.5$ and $p=1$ respectively.

For the fully parallel dynamics $p=1$, we get
$V(s)=2/(2+\rho s)$ and hence
$Q(s)=4/(2+\rho s)^2$. By inverting the Laplace transform, we recover the
result $P(m)=\frac{4 m}{\rho^2} e^{-2 m/\rho}$ obtained by Coppersmith et.
al\cite{SC}. In the random sequential limit, $p\to 0$, we get 
\bea
V(s)&=&\frac{-1+\sqrt{1+2\rho s}}{\rho s} \nonumber \\
Q(s)&=&\frac{1}{\sqrt{1+2 \rho s}} \qquad and\nonumber \\
P(m)&=&\frac{1}{\sqrt{2 \pi \rho m}} \exp(-m/(2 \rho)).
\label{ten}
\eea
The same result was obtained by Krug and Garcia\cite{krug} by a somewhat
indirect method by computing the moments and then guessing the
distribution from its moments. When $p=0.5$, the expression for $P(m)$ is
complicated and we do not reproduce it here.

For arbitrary $p$, a closed form expression of $P(m)$ is difficult to
obtain. However the asymptotic behaviour of $P(m)$ for large and small $m$
can be easily derived. For large $m$, we expect $P(m)\sim e^{-\alpha m}$.
The decay coefficient $\alpha$ can be derived by noting that the Laplace
transform $Q(s)$ must have a pole at $s=-\alpha$. From Eq. (\ref{eight}), we 
note
that the pole of $Q$ occurs when $V=(1-q^2)/pq$. Using this in 
Eq. (\ref{nine}),
we get,
\bea
\alpha= {{2(1-p)^{1-p}}\over {\rho (2-p)^{2-p}} }.
\label{eleven}
\eea
In the limits, $p=1$ and $p\to 0$, this gives the correct decay
coefficient of $P(m)$.

For small $m$, on the other hand, $P(m)$ has an algebraic tail, $P(m)\sim
m^{-\beta}$. From Eq. (\ref{nine}), we note
that for large $s$, $V(s)\approx (2/\rho s)^{1/(2-p)}$. Using
$Q(s)=d(sV)/ds$, we get for large $s$,
\bea
Q(s)\approx {{1-p}\over {2-p}}({2\over {\rho s}})^{1/(2-p)}.
\label{twelve}
\eea
This implies that for $p<1$, $P(m)\sim m^{-\beta}$ for small $m$ with
$\beta=(1-p)/(2-p)$. Note that for $p=1$, the coefficient of $1/s$
vanishes in Eq. (\ref{twelve}) and the leading order term decays as $1/s^2$,
implying $\beta=-1$ for $p=1$. Thus there is a discontinuity in the
exponent $\beta(p)$ at $p=1$.

How good is the product measure ansatz? In general, we have noted before
that the equation of the n-point 
distribution function contains the (n+1)-point distribution function.
If the product measure ansatz were to
be exact, then one has to ensure that every equation of the hierarchy
is satisfied by the product measure ansatz. This was in fact proved
to be case for $p=1$\cite{SC}. It is easy to show that this
ansatz is exact only for $p=1$ and fails for all $p<1$. This is proved
by showing that for $p<1$, the second equation of the hierarchy (involving
the two-point and three-point distributions) is not satisfied by
the $P(m)$ obtained from the first equation of the hierarchy, i.e., Eq.
(\ref{seven}) assuming product measure. Algebraic details are given in
Appendix A.

For $p<1$, we compared the mean field 
answer for $P(m)$ with the numerically obtained distribution in one
dimension. In the limit $p\rightarrow
0$, the mean field $P(m)$
matches extremely well with the numerically computed one.
This was also noted by Krug and Garcia\cite{krug}.
For arbitrary $p$, we do not have a closed form expression of mean
field $P(m)$ to compare with the simulation results. However, various 
moments of $m$ with the mean field $P(m)$ can be computed exactly for
arbitrary $p$ and compared to the numerically obtained moments.
The mean field moments are computed by expanding $V(s)$
in
powers of $s$. We list the the moments $<m^n>$ for $n=1,\ldots 5$ below.
\bea
<m>&=&\rho \nonumber \\
<m^2>&=&\frac{3 (2-p)}{2}\rho^2 \nonumber \\
<m^3>&=&\frac{3 (2-p) (5-3 p)}{2}\rho^3 
\label{thirteen} \\
<m^4>&=&\frac{5 (2-p) (21-26 p+8 p^2)}{2}\rho^4 \nonumber \\
<m^5>&=&\frac{15 (2-p) (504-955 p+600 p^2-125 p^3)}{16}\rho^5 \nonumber 
\eea
To check how accurate these mean field moments are, we have
computed these moments directly from numerical simulation of the
model for different values of $p$. In Table I, we compare the mean field
moments (up to order $5$) to the numerical ones for a representative
value of $p=0.8$.
The closeness of these moments to the numerical values for arbitrary $p$
suggests strongly that the mean field $P(m)$ may be exact for all $p$ even
though the product measure fails for $p<1$.

\subsection{Correlation Function}

In this subsection we compute the two point mass correlation function
exactly for the asymmetric continuous
mass model. In the continuous time case ($p\to 0$ limit of our model),
this was computed exactly by Krug and Garcia\cite{krug} for arbitrary 
probability distributions of the random fraction $r$. We reproduce their 
calculation here for completeness.
Multiplying $m_i(t+1)$ by $m_j(t+1)$ in Eq. (\ref{five}) and taking 
expectation value in the steady state, we find that two point correlations  
$C_j=<m_i m_{i+j}>$ satisfy the following set of linear equations, 
\bea
C_0 (\mu_1-\mu_2) - C_1 \mu_1 (1-\mu_1) &=&0,\nonumber \\
C_0 (\mu_1-\mu_2) - 2 C_1 \mu_1 (1-\mu_1) +C_2 \mu_1 (1-\mu_1)  &=&0, 
\label{fourteen}\\
C_{j-1} - 2 C_j+C_{j+1} &=&0,\qquad j=2,3,4,\ldots \nonumber 
\eea
with the boundary conditions $C_j \to \rho^2$ as $j\to \infty$. Here
$\mu_1=\langle r_i\rangle$ and $\mu_2=\langle {r_i}^2\rangle$ are the
first and second moments of the random fraction $r_i$ distributed 
according to Eq. (\ref{six}). These set of of equations can be solved easily
to give\cite{krug},
\bea
C_0&=&\frac{\mu_1(1-\mu_1)}{\mu_1-\mu_2}\rho^2 \nonumber \\
C_j&=&\rho^2, \qquad j=1,2,3,\ldots
\label{fifteen}
\eea
Thus the two point correlation function $<m_i m_j>$ is equal to $<m_i><m_j>$
for $i\neq j$. In fact this conclusion holds for any arbitrary
distribution of the fractions $r_i$. This however does not mean
that the product measure is exact. That would require that all higher
order correlations must also factorize. In fact, for the asymmetric 
model, it can be shown\cite{RM} that the 3-point correlation function does
not factorize except for $p=1$.

\section{The Symmetric Model}

In this section, we study the continuous mass model where mass
transport has no bias in direction. Once again we have a continuous mass
$m_i$ at each site. In each time step, a fraction is chosen at random from
a uniform distribution on $[0,1]$ and this fractional mass is transported
to the left or right nearest neighbour with equal probability $p/2$.
With probability $q=1-p$, the fractional mass stays at the original
site.
In order to write down the mass evolution equation, it is convenient to
define a set of variables $s_i$ at each site $i$. The variable $s_i$ can
be either $+1$ or $-1$ with equal probability $1/2$. If $s_i=1$, it
indicates that the fractional mass from site $i$ is transported to the
right neighbour. On the other hand, $s_i=-1$ indicates transport to
the left neighbour. Then the mass evolution equation can be written
down as in the asymmetric case,
\bea
m_i(t+1)=(1-r_i)m_i(t)+{{1+s_{i-1}}\over
{2}}r_{i-1}m_{i-1}(t)+{{1-s_{i+1}}\over {2}}r_{i+1}m_{i+1},
\label{sixteen}
\eea 
where the random variables $r_i$ have the same distribution
$f(r_i)=q\delta(r_i)+p$ as in the asymmetric case. The evolution of the
single point mass distribution function $P(m,t)$ can be written down as in
the asymmetric case (Eq. (\ref{seven})). The only difference is that now the
single point equation contains three point distribution (as opposed to
the two point function in the asymmetric case) and the additional $s_i$
variables. 

\subsection{Mean Field Theory}
Assuming product measure, this equation can be solved
in the same fashion as in the asymmetric case. It follows that the
Laplace transform $Q(s)$ of $P(m)$ in the stationary state satisfies,
\bea
Q(s)={{pV(q+pV+1)^2}\over {4-q(q+pV+1)^2}}
\label{seventeen}
\eea
where $V(s)=\int_0^{\infty}Q(su)du$ as in the asymmetric case.
Using $Q(s)=d(sV)/ds$, we find the function $V(s)$ is given by the
solution of the following nonlinear equation,
\bea
[1-{{p(1-V)}\over {4}}]^{\frac{p}{4-p}}
V^{-\frac{8-5p+p^2}{4-p}}(1-V)={{\rho s}\over {2}}.
\label{eighteen}
\eea

In the limit $p\rightarrow 0$ (random sequential limit), this equation can
be solved in closed form and we get,
\bea
P(m)=\frac{1}{\sqrt{2 \pi \rho m}} \exp(-m/(2 \rho)),
\label{nineteen}
\eea
which has the same expression as for the asymmetric $p\to 0$ case. For
other values of $p$, while we are unable to get a closed form expression,
the asymptotic behaviour of $P(m)$ for large and small $m$ can be
easily derived. We find that for large $m$, $P(m)\sim
\exp(-\alpha m)$ where the coefficient $\alpha(p)$ can be determined
in the same way as in the asymmetric case. It is given by a long expression
which we do not present here. For small $m$, $P(m)$ has an algebraic tail,
$P(m)\sim m^{-\beta}$ where the exponent $\beta (p)$ can be determined 
by examining the large $s$ behaviour of $Q(s)$. We
find $\beta (p)=(2-p)^2/(8-5p+p^2)$ which decreases continuously from
$1/2$ ($p\to 0$) to $1/4$ ($p=1$).

For the symmetric case, we show in Appendix-B that the product measure is
exact in the $p\to 0$ limit. For $p>0$, the product measure fails
and unlike the asymmetric case, the mean field $P(m)$ is considerably
different from the distribution obtained numerically. This failure
of mean field theory for $p>0$ shows up in the calculation of two point
correlation function as done in the next subsection. However, while the mean
field theory fails for large $m$ (as evident from expectation value of
the moments of the mass distribution shown in Table II), it matches very 
well with the numerical result for small $m$ (see figure 3).

\subsection {Correlation Function}

For the symmetric model, the translationally invariant stationary two
point mass correlation
function, $C_{j-i}=\langle m_im_j\rangle$ does not factorize for $j\neq i$
as in the asymmetric case. Below we compute the two point correlation
exactly and show that the connected part of
the correlation function in fact has a nontrivial spatial dependence. 

Multiplying Eq. (\ref{sixteen}) by $m_j(t+1)$ and taking expectation value, 
we find
that in the stationary limit $t\to \infty$, the correlation function $C_j$
satisfies,
\bea
\mu_1 C_{j-2} (1-\delta_{j2})+4 (1-\mu_1)C_{j-1}+2 (3 \mu_1
-4)C_j+4(1-\mu_1)C_{j+1}+\mu_1C_{j+2}&=&0 \quad j=2,3,\ldots \nonumber\\
4(1-w) C_{0} +2 (7\mu_1/2 -4) C_1+4 (1-\mu_1) C_2 +\mu_1 C_3 &=&0
\label{twenty}\\
4(-1+w) C_{0} +4 (1-\mu_1) C_1 +\mu_1 C_2 &=&0, \nonumber
\eea
where $\mu_1=\langle r_i\rangle$ and $\mu_2=\langle {r_i}^2\rangle$ are
respectively the first and second moments of $f(r_i)$ and $w=\mu_2/\mu_1$.

Let $G(z)=\sum_{j=1}^{\infty}C_j z^j$ be the generating function.
Multiplying Eq. (\ref{twenty}) by $z^j$ and summing over $j$'s, we get 
\bea
G(z)=\frac{z[4(1-w)z C_0 +C_1 \mu_1 (1+z)]}{(1-z)[4 z+\mu_1 (1-z)^2]}
\label{twentyone}
\eea
The boundary condition, $C_j\to \rho^2$ as $j\to \infty$ implies that
$G(z)\rightarrow \rho^2/(1-z)$ as $z\to 1$. This
gives us one relation between $C_0$ and $C_1$, 
\bea
C_1=(2 \rho^2-2(1-w)C_0)/\mu_1.
\label{twentytwo}
\eea
We need one more condition to fix both $C_0$ and $C_1$. This is obtained
by noting that $G(z)$ in Eq. (\ref{twentyone}) has three poles, $z=1$ and 
$z=z_{\pm}$
where $z_{\pm}= (\mu_1-2\pm 2\sqrt{1-\mu_1})/\mu_1$. We note that
$|z_{+}|<1$ which would imply that $C_j$ will blow up exponentially
as $|z_{+}|^j$ for large $j$. Since this can not happen, the numerator
on the right hand side of Eq. (\ref{twentyone}) must also vanish at 
$z=z_{+}$ in order
to cancel the pole. This provides an additional condition which together 
with Eq. (\ref{twentytwo}) gives,
\bea
C_0=\frac{\rho^2 (1+z_{+})}{(1-w)(1-z_{+})}=\frac{\rho^2 \sqrt
{1-\mu_1}}{1-w}
\label{twentythree}
\eea
and $C_1$ can be determined from Eq. (\ref{twentytwo}). Inverting the 
generating
function, we find that for any $n>0$, 
\bea
C_n=\rho^2[1-z_{+}^n].
\label{twentyfour}
\eea
Since $z_{+}=(\mu_1-2+ 2\sqrt{1-\mu_1})/\mu_1$ lies in $[-1,0]$,
clearly the connected part of the correlation function has a nontrivial
oscillation with
distance with an amplitude that decays exponentially with the distance.

Curiously the function $C_n$ for $n>0$ depends only on $\mu_1$ but not on
$\mu_2$, whereas $C_0$ involves both $\mu_1$ and $\mu_2$. We also note
that the above exact result is valid for any arbitrary distribution $f(r)$
of the fractions $r_i$ and not just for the special distribution
given by Eq. (\ref{six}). For that distribution, we get from Eq. (\ref{six}), 
$\mu_1=p/2$
and $\mu_2=p/3$ and hence $w=2/3$. One useful check is that in the
limit $p\to 0$, we get $z_{+}\to 0$ implying complete decoupling of
the two point correlation. This is consistent with the fact that
product measure is exact in the symmetric case only in the $p\to 0$ limit.
For $p>0$, the correlation function has a nontrivial spatial dependence
and product measure clearly fails.
 
\section{The Discrete Mass Model}
In this section we study the model when the mass $m_i$ at each site $i$ 
is a discrete non negative integer. In each time step, a block
of size $n_i$ is chosen at each site and is transported to its
neighbour with probability $p$ and stays at the original site with
probability $q=1-p$. The block size $n_i$ is a discrete random variable
which can take values $0$, $1$, $2, \ldots$, $m_i$, all with equal
probability $1/(m_i+1)$. As in the continuous mass model, the mass
transport can be either asymmetric or symmetric. We study here only the
asymmetric model but the symmetric version can be studied by using
similar procedures.

There is an equivalent lattice gas representation of this model in one
dimension as mentioned in Section-(II). In this mapping,
lattice site $i$ of the mass model corresponds to the $i$-th
hard core particle and the mass $m_i$ represents the number of holes or
empty sides between the $i$-th and $(i+1)$-th particle. In the lattice
gas dynamics of the asymmetric model, at each time step every particle
jumps to any one of the
available vacant sites in front of it with equal probability. 

The analysis of the stationary mass distribution of the asymmetric
discrete model proceeds along the same line as its continuous 
counterpart. We write down the evolution equation of the
single site distribution function $P(m,t)$ in terms of the joint
two point distribution $P(m_1,m_2,t)$. Assuming product measure holds,
the evolution equation is given by,
\bea
P(m_i,t+1)&=&p^2 \sum_{m_{i-1}=0}^{\infty}\sum_{m_1=0}^{m_{i-1}}\sum_{m_i^\p=0}^
{\infty} \sum_{m_2=0}^{m_i^\p}
\frac{P(m_{i-1})P(m_i^\p)}{(m_{i-1}+1)(m_i^\p+1)}\delta(m_i^\p
-m_2+m_1-m_i)\nonumber\\
 &+ &p q \sum_{m_i^\p=0}^ {\infty} \sum_{m_2=0}^{m_i^\p}
\frac{P(m_i^\p)}{(m_i^\p+1)}\delta(m_i^\p -m_2-m_i)\nonumber\\
&+ &pq \sum_{m_{i-1}=0}^{\infty}\sum_{m_1=0}^{m_{i-1}}\sum_{m_i^\p=0}^
{\infty} \frac{P(m_{i-1})P(m_i^\p)}{(m_{i-1}+1)}\delta(m_i^\p
+m_1-m_i)\nonumber\\
&+ &q^2 \sum_{m_i^\p=0}^ {\infty} P(m_i^\p)\delta(m_i^\p-m_i).
\label{twentyfive}
\eea
We define the generating function, $Q(x)=\sum_0^{\infty}P(m)x^m$. In the
stationary limit, we get from the above equation,
\be
Q(x)=\frac{[f(x)-f(1)][p(f(x)-f(1))+q(x-1)]}{(x-1)[(1+q)(x-1)-q(f(x)-f(1))]}
\label{twentysix}
\ee
where $f(x)=\sum_{m=0}^{\infty}\frac{P(m)}{m+1} x^{m+1}$.
Using $Q(x)=\frac{df}{dx}$, one can obtain closed form expressions of
$Q(x)$ and hence of $P(m)$ only in the two limits, $p=1$ and $p\to 0$.
For the fully parallel dynamics ($p=1$), we get,
\bea
Q(x)&=&\frac{1}{(1-\frac{\rho}{2} (x-1))^2}\\
\label{twentyseven}
P(m)&=&\frac{4(m+1)\rho^n}{(\rho+2)^{n+2}}.
\label{twentyeight}
\eea
For the random sequential case ($p\rightarrow 0$), we find
\bea
Q(x)&=&\frac{1}{\sqrt{1-2\rho(x-1)}}\\
\label{twentynine}
P(m)&=&\frac{(2\rho)^m}{(1+2 \rho)^{n+1/2}}\frac{(2 n)!}{(n!)^2 2^{2n}}.
\label{thirty}
\eea
It is easy to check that in the limit of large $m$ and $\rho$, these
distributions reduce to their continuous counterparts (Eq.
(\ref{two}) and Eq. (\ref{three}) respectively) as expected.

As in the continuous asymmetric model, it turns out that the product
measure is exact only in the $p=1$ limit. The proof that the product
measure is exact for $p=1$ in the discrete case can be derived by
following the same line of arguments as used for the continuous
case\cite{SC}. Basically, one writes down the exact evolution
equation for the $n$-point joint distribution which involves the
$(n+1)$-th point joint distribution. One makes the ansatz for product
measure and ensures that this ansatz is consistent for all values of $n$,
i.e., all the equations of the hierarchy satisfy the product measure
ansatz. 

Without giving the details we just outline below few basic steps.
For $p=1$, assuming product measure in the equation involving single point
and two point distributions (Eq. \ref{twentyfive}), we obtain $P(m)$ as 
given by Eq. \ref{twentyeight}.
Consider first a cluster of $n$ neighbouring sites
$1,2,..,n$. The time development for the $n$- point probability
distribution can be written as
\bea
P(m_1,\ldots,m_{n},t+1)&=&\sum_{m_{0}^\p=0}^{\infty}\sum_{r_0=0}^{m_0^\p}
\ldots \sum_{m_{n}^\p=0}^ {\infty} \sum_{r_{n}=0}^{m_{n}^\p}
\frac{P(m_{0}^\p,m_1^\p,\ldots,m_{n}^\p,t)}{(m_{0}^\p+1)(m_1^\p+1)\ldots
(m_n^\p+1)}\nonumber\\
& &\delta(m_1^\p -r_1+r_{0}-m_1)\ldots \delta(m_{n}^\p
-r_{n}+r_{n-1}-m_n).
\label{thirtyone}
\eea
We have to now show that the product measure ansatz
$P(m_1,m_2,\ldots)=\prod_i P(m_i)$ in the steady state with $P(m)$ given
by Eq. (\ref{twentyeight}) is consistent with Eq. (\ref{thirtyone}). To
show this, we
consider the n-variable generating function
$Q(x_1,\ldots,x_n)=\sum_{m_1=0}^{\infty}\ldots \sum_{m_n=0}^{\infty}
P(m_1,\ldots , m_n) x_1^{m_1} \ldots x_n^{m_n}$. We assume product
measure on the right hand side of Eq. (\ref{thirtytwo}), sum over the $m_i$'s 
and obtain,
\bea
Q(x_1,\ldots,x_n)&=&\frac{\alpha-f(x_1)}{1-x_1}
\frac{f(x_2)-f(x_1)}{x_2-x_1}
\ldots \frac{f(x_n)-f(x_{n-1})}{x_n-x_{n-1}}
\frac{\alpha-f(x_n)}{1-x_n}\nonumber\\
&=& Q(x_1) Q(x_2) \ldots Q(x_n)
\label{thirtytwo}
\eea
where in deriving the last step we have used $Q(x)=df/dx$ and the
expression of $Q(x)$ from Eq. (\ref{twentysix}). One can repeat the 
same calculation
when the $n$ sites are not necessarily neighbours. This therefore proves
that every equation of the hierarchy of distribution functions satisfies
the product measure ansatz for $p=1$. However, this proof fails for $p<1$
as in the continuous case and the same line of argument used for the
continuous case (see Appendix A) goes through for the discrete case. Even 
though the product measure fails, the mean field answer for other values of
$p$ match very well with the numerically computed one. For the random
sequential case ($p\rightarrow 0$), we compare 
the mean field result for the single site probability distribution with 
numerical simulation (see figure 4).

\section{Summary and Conclusion}

In this paper we have studied a simple mass model of chipping and
aggregation
where a mass at a site can chip off a fraction to its neighbour. A
parameter $p$ was introduced which allowed us to interpolate between
parallel
dynamics and random sequential dynamics. 
We studied the model for both
continuous and discrete masses as well as for symmetric and asymmetric
transport of mass. 

We have calculated analytically the mass distribution function $P(m)$ in
the steady state for all $p$ assuming product measure, i.e., neglecting
correlation between masses. In some cases we proved that this product
measure is exact. One of the main results is that the distribution
$P(m)$ has an algebraic tail for small $m$, $P(m)\sim m^{-\beta(p)}$
where the exponent $\beta(p)$ depends on $p$. Thus the steady state is
non universal and depends on the specific nature of the dynamics
characterized by the parameter $p$.

Another interesting point is that for the asymmetric continuous mass
model, we show that even though the two-point mass correlation function
decouples for any $p$, product measure is not valid for $p<1$. 
This means
that the correlations between masses at different sites show up only
in $3$ or higher order correlation functions but not at the $2$ point
level. Exact calculation of the $3$-point correlation function will be
presented elsewhere\cite{RM}. Interestingly however the single point mass
distribution $P(m)$ obtained using product measure ansatz is extremely
close to the numerically obtained distribution.

Interpreting $m_i$ as the distance between two hard core particles
labelled $i$ and $(i+1)$, it is easy to see that within product measure
ansatz, the steady state probability of a given configuration can be
written as, $P(m_1,m_2,\ldots)\sim \prod_i {m_i}^{-\beta(p)}$ for small
gaps between neighbouring particles. This represents a gas of
particles moving on a ring
with nearest neighbour interaction $\beta(p) \log(r)$ for small $r$, where
$r$ is the separation between neighbouring particles. 
Choosing different dynamics via tuning $p$ corresponds to changing the
coupling continuously. 
For the asymmetric model, $\beta(p)=(1-p)/(2-p)$ for $p<1$ and $-1$ for
$p=1$. This corresponds to a shift from a potential that
prefers "bunching" of particles for $p<1$ to a repulsive one at $p=1$. This
jump discontinuity is lost for the symmetric model where we have
$\beta(p)= (2-p)^2/(8-5p+p^2)>0$ for all $p$.

We also calculated exactly the correlation function $C_j=<m_0 m_j>$ for
the asymmetric and the symmetric models. When the
transport is asymmetric the correlation function factorises for $j\neq 0$.
Unlike the asymmetric case, there are nontrivial correlations in
the symmetric model. The connected part of the correlation function oscillates
with distance and the amplitude of the oscillation decays exponentially
with distance (see figure 5). 

A simple lattice mass model with diffusion, aggregation and chipping of
single units of mass was shown to exhibit nonequilibrium phase transition
in the steady state\cite{MKB,MKB1,Krapiv}. In this paper we have shown
that if a random fraction chips off instead of a single unit,
the steady state no longer has a phase transition as the rates of
microscopic processes are varied.

There are several directions for future work.
For the asymmetric continuous mass model with continuous dynamics ($p\to
0$ limit of our model), Krug and Garcia\cite{krug} had derived
density-density correlations between particles in the lattice gas
representation. It would be interesting to extend their calculation to
general $p$. Another interesting direction would be to derive
the large scale hydrodynamics for general $p$ and extend the calculation
of the tracer diffusion coefficient\cite{krug,GS} to general $p$.

We thank D. Dhar, M. Barma and J. Krug for very useful discussions.

\section*{Appendix A: Non Exactness of product measure ansatz for
asymmetric continuous model for \q $<1$}
In this appendix, we show that the product measure or the mean field
theory is not exact for
asymmetric continuous mass model for $p\neq 1$. The steps in the proof are
as follows. In general the evolution equation of the $n$-point joint
distribution function will involve the $(n+1)$-point distribution.
If the product measure were to be exact, then every equation of this
hierarchy has to be consistent with that ansatz. We show below that
for the asymmetric case, the second equation of the hierarchy namely 
the one involving the $2$-point and $3$-point distributions is not
consistent with product measure ansatz.

Firstly, we recall that we can derive an expression for the single point
distribution $P(m)$ in the steady state by assuming product measure in
the equation involving the single point and two point distributions
(namely Eq. (\ref{seven})). The Laplace transform $Q(s)=pV(pV+q)/(1-pqV-q^2)$
is given by Eq. (\ref{eight}) where $Q(s)=d(sV)/ds$. 
Next we write down the
second equation of the hierarchy, namely the evolution equation of the
joint mass distribution, $P(m_i,m_{i+1})$ of two adjacent sites $i$
and $(i+1)$,
\bea
P(m_i,m_{i+1},t+1) & = &\int dm_{i-1}\int dr_{i-1}f(r_{i-1}) 
\int dm_i^\p\int dr_i f(r_i)
\int dm_{i+1}^{\p}\int dr_{i+1}f(r_{i+1}) \nonumber \\
&\times&P(m_{i-1},m_{i},m_{i+1},t)\delta
(m_{i-1}r_{i-1}+m_i^\p(1-r_i)-m_{i})\nonumber\\
&\times&\delta (m_{i}^\p r_{i}+m_{i+1}^\p(1-r_{i+1})-m_{i+1}).
\label{thirtythree}
\eea
All the integrals over $dm$ run from $0$ to $\infty$ while the integrals over
$dr$ run from $0$ to $1$. $ P(m_{i-1},m_{i},m_{i+1},t)$ is the three point
joint mass distribution function and $f(r)$ is given by Eq. (\ref{six}). If the
product measure were exact, the
joint distributions in the above equation would factorize and the
resulting equation must be satisfied by the $P(m)$ 
obtained from the first equation of the hierarchy, namely Eq.
(\ref{eight}). 

Assuming factorization $P(m_1,m_2\ldots)=\prod_i P(m_i)$ in Eq.
(\ref{thirtythree}), multiplying both sides by 
$e^{-m_i s_1-m_{i+1}s_2}$ and then integrating over $m_i$ and $m_{i+1}$,
we get
\bea
Q(s_1)Q(s_2)=\left(q+pV(s_1)\right)
\left(qQ(s_2)+pV(s_2)\right)
\left(qQ(s_1)+p\int_0^{1}dr_iQ(s_1+(s_2-s_2)r_i)\right).
\label{thirtyfour}
\eea
If product measure were to be true, this equation must be satisfied with
the expression of $Q(s)$ obtained from Eq. (\ref{eight}).
If we substitute the expression for $Q(s)$ from Eq. (\ref{eight}) in the above
equation, we find after a somewhat tedious but straightforward algebra,
that Eq. (\ref{thirtyfour}) reduces to,
\be
\left(\frac{V(s_2)}{V(s_1)}\right)^{2-p}-(2-p)\frac{V(s_2)}{V(s_1)}+1-p=0.
\label{thirtyfive}
\ee
If product measure is to be
true then a necessary condition (but not sufficient) is that the above
equation be
satisfied for arbitrary values of $s_1$ and $s_2$.
For $p<1$, this is an algebraic equation for the ratio
$V(s_2)/V(s_1)$. Since the coefficients do not involve $s_1$ or $s_2$, the
solution for $V(s_2)/V(s_1)$ will be a constant independent of $s_1$ and
$s_2$. Clearly this can not be true for arbitrary values of $s_1$ and
$s_2$. Thus product ansatz is not exact 
for $p<1$ for asymmetric dynamics.

Note however that for $p=1$, Eq. (\ref{thirtyfive}) becomes an identity.
This
however is a necessary but not sufficient condition to prove that product
measure is exact for $p=1$. However it was shown\cite{SC} that for $p=1$,
all equations of the hierarchy of distribution functions are actually 
consistent with product measure ansatz.

\section*{Appendix B: Proof of exactness of product ansatz for
\q$\rightarrow 0$ for the symmetric model}
In this appendix we show that the mean field is exact for the $p\rightarrow
0$ limit of the symmetric continuous model. 
Consider a cluster of $n$ consecutive sites $1,2,\ldots,n$. In the steady
state, the joint probability distribution
function $P(m_1,m_2,\ldots,m_n)$ satisfies the equation,
\bea
0&=&-(2 n+2) P(m_1,\ldots,m_n)\nonumber \\
&+&\li dm_0\ri dr \li dm_{1}^{\p}P(m_0,\ldots,m_n) 
(\delta (m_1^{\p}+m_0 r-m_1) 
+\delta (m_1^{\p}(1-r)-m_1)) 
\nonumber \\
&+&\li dm_{n+1}\ri dr \li dm_{n}^{\p}P(m_1,\ldots,m_{n+1}) 
(\delta (m_n^{\p}+m_{n+1} r-m_n) 
+\delta (m_n^{\p}(1-r)-m_n))
\nonumber \\
&+& \sum_{i=1}^{n-1} \li dm_i^{\p}\ri dr \li dm_{i+1}^{\p} P(m_1,\ldots,m_n)
\nonumber \\
&\times&(\delta (m_i^{\p} (1-r)-m_i)\delta(m_{i+1}^{\p}+m_i^{\p}
r-m_{i+1})+
\delta (m_i^{\p}+m_{i+1}^{\p}r-m_i)\delta(m_{i+1}^{\p}(1-r)-m_{i+1})).
\label{thirtysix}
\eea
The first term is the total rate of going out of the state. The second and
third terms describe the mass transfer at the boundary of the n-cluster while
the last term accounts for mass transfer within the cluster.
Let $P(m_1,\ldots,m_n)=\prod_{1}^{n} P(m_i)$. We multiply both sides of
the equation by $e^{-m_1 s_1-\ldots-m_n s_n}$ and sum over
$m_1,\ldots,m_n$. The
resulting terms in the right hand side can be simplified by using
the explicit expression of $V(s)$ from Eq. (\ref{ten}).
Then each one of the terms involving the integrals 
reduces to 
$2 \prod_{1}^{n}Q(s_i)$, where $Q(s)=\int_{0}^{\infty}P(m) e^{-m s}dm$ as
before. Thus Eq. (\ref{thirtysix}) is indeed satisfied by the 
product measure ansatz for all $n$. Joint
probability distributions for any $n$ arbitrary sites can be split up into
product of distributions for clusters of neighbouring sites, and then the 
proof can be applied for each of the individual clusters. 

We note that for the $p\to 0$ limit of the symmetric model, it was shown in 
Ref.\cite{krug} by a different method that the product measure is exact for 
any finite system of size $N$.

For symmetric model with $p>0$, the product measure is not exact as
was shown in the text by explicit calculation of two point
mass correlation function.

\newpage
\begin{figure}
\begin{center}
\leavevmode
\psfig{figure=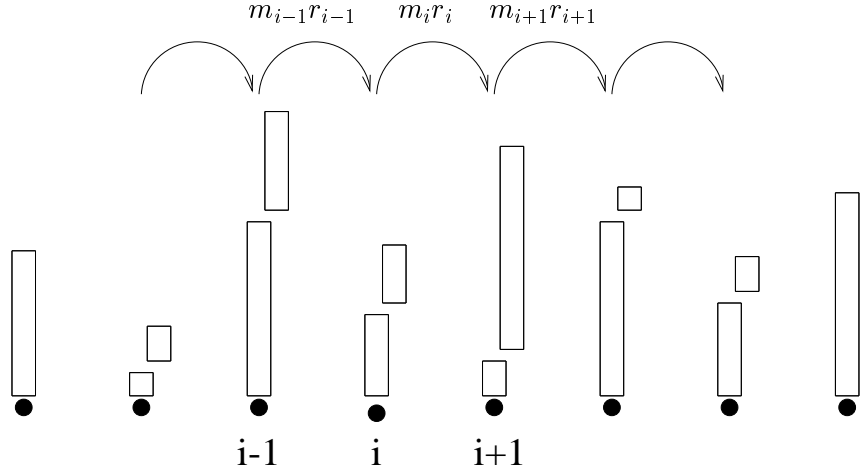,width=12cm,angle=0}
\caption{Asymmetric continuous mass model: A random fraction $r_i$ of each
mass $m_i$ is broken off and
added to the right neighbour with probability $p$. With probability
$(1-p)$, the broken piece rejoins the original mass.}
\label{fig:model_def}
\end{center}
\end{figure}

\begin{figure}
\begin{center}
\leavevmode
\psfig{figure=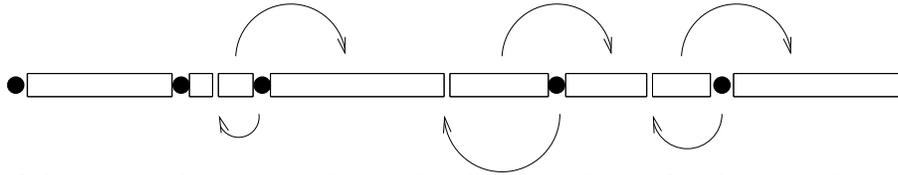,width=12cm,angle=0}
\caption{Mapping of the mass model to a particle model is shown. Each
transfer of mass to the right corresponds to the particle (filled circles)
jumping to the
left.}
\label{fig:mapping_particle}
\end{center}
\end{figure}

\begin{figure}
\begin{center}
\leavevmode
\psfig{figure=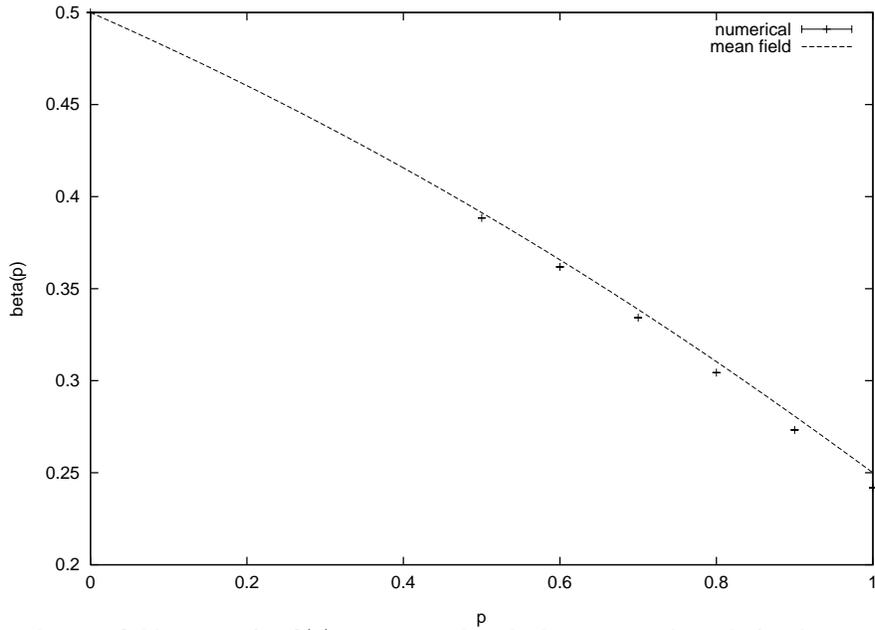,width=12cm,angle=-90}
\caption{The analytical mean field answer for $\beta(p)$ is compared with the
numerical result for the symmetric continuous mass model. While the numerical
single site distribution for masses is quite different from the mean field
answer (see Table II), the small $m$ behaviour is predicted well by the 
mean field.}
\label{fig:ran_seq_comp}
\end{center}
\end{figure}

\begin{figure}
\begin{center}
\leavevmode
\psfig{figure=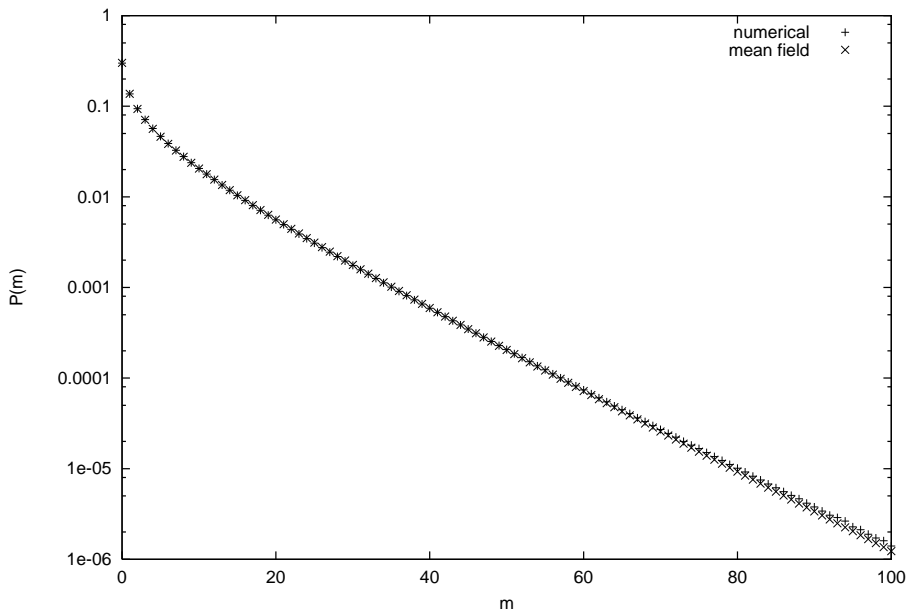,width=12cm,angle=-90}
\caption{The analytical mean field answer for the single site mass
distribution function $P(m)$ given by Eq. (\ref{thirty}) is compared to
the numerical result for the 
asymmetric random sequential {\em {discrete}} mass model. The data is for
system size L=20000. The closeness of the two curves suggest that
mean field $P(m)$ is exact even though the product measure fails.}
\label{fig:ran_seq_disc}
\end{center}
\end{figure}

\begin{figure}
\begin{center}
\leavevmode
\psfig{figure=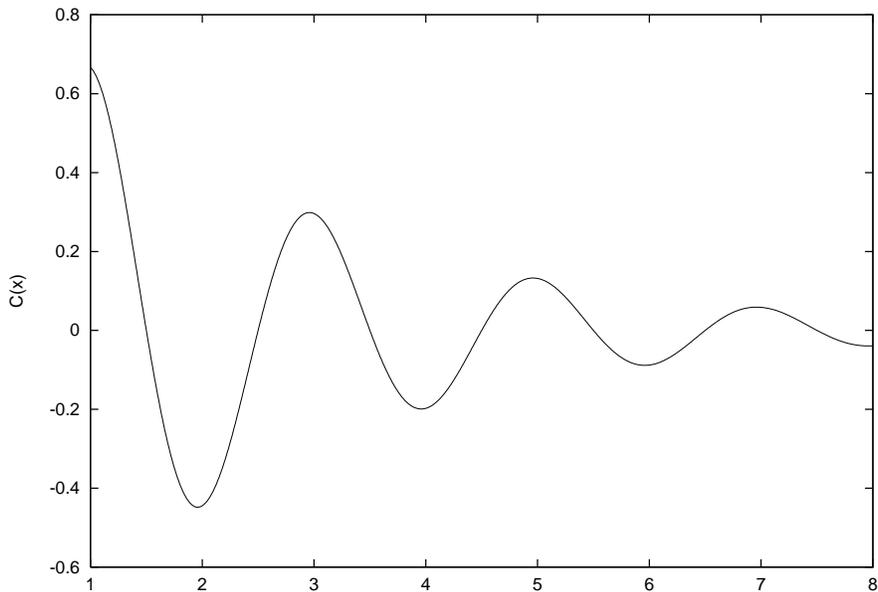,width=12cm,angle=-90}
\caption{The correlation function $C(x)=<m_0 m_x>-<m>^2$ given by Eq.
(\ref{twentyfour}) is shown for $\mu_1=0.96$.
It oscillates with distance with the amplitude decaying 
exponentially to zero.}
\label{fig:transfer}
\end{center}
\end{figure}

\begin{table}
\caption{
Numerically obtained moments of the mass are compared with
the mean field values (Eq. \ref{thirteen}) for $p=0.8$ in the asymmetric
continuous model. The excellent agreement with the mean field results in
this case is to be contrasted
with rather poor
agreement with mean field results in case of  symmetric continuous model
(see Table II).}
\begin{tabular}{|c|l|c|} 
Moments & Numerical & Mean Field \\ \hline
$<m^2>$ & 1.7998(0) & 1.80       \\ \hline
$<m^3>$ & 4.6826(0)& 4.68 \\ \hline
$<m^4>$ & 15.9888(3)& 15.96 \\ \hline
$<m^5>$ & 67.72(1)& 67.50 \\ \hline
\end{tabular} 
\end{table}

\begin{table}
\caption{
Comparison of numerically obtained moments of the mass with
the mean field values for $p=0.8$ in the symmetric
continuous model. This clearly shows that the mean field approximation
is not good for the symmetric case as compared to the asymmetric case. }
\begin{tabular}{|c|l|c|} 
Moments & Numerical & Mean Field \\ \hline
$<m^2>$ & 2.3237(4) & 2.100       \\ \hline
$<m^3>$ & 8.623(5)& 6.660 \\ \hline
$<m^4>$ & 44.37(7)& 28.260 \\ \hline
$<m^5>$ & 293.2(9)& 150.314 \\ \hline
\end{tabular} 
\end{table}


\begin{thebibliography}{999}

\bibitem{White} W. H. White, J. Colloid Interface Sci. {\bf 87}, 204
(1982).

\bibitem{Ziff} R. M. Ziff, J. Stat. Phys. {\bf 23}, 241 (1980).

\bibitem{Krapiv} P. L. Krapivsky and S. Redner, Phys. Rev. E {\bf 54},
3553 (1996).

\bibitem{river} A. E. Scheidegger, Bull. I.A.S.H. {\bf 12}, 15 (1967).

\bibitem{Fried} S. K. Friedlander, {\it Smoke, Dust and Haze} (Wiley
Interscience, New York, 1977).

\bibitem{Lewis} B. Lewis and J. C. Anderson, {\it {Nucleation and Growth
of Thin Films}} (Academic, New York, 1978).

\bibitem{Tak} H. Takayasu, Phys. Rev. Lett. {\bf 63}, 2563 (1989); H.
Takayasu, I. Nishikawa and H. Tasaki, Phys. Rev. A {\bf 37}, 3110 (1988).

\bibitem{MKB} S. N. Majumdar, S. Krishnamurthy and M. Barma, Phys. Rev.
Lett. {\bf 81}, 3691 (1998).

\bibitem{MKB1} S. N. Majumdar, S. Krishnamurthy and M. Barma,
cond-mat/9908443; to appear in J. Stat. Phys.

\bibitem{derrida} B. Derrida, M. R. Evans, V. Hakim and V. Pasquier, J.
Phys. A. {\bf 26}, 1493 (1993).

\bibitem{dhar} D. Dhar, Phys. Rev. Lett. {\bf 64}, 1613 (1990); D. Dhar
and R. Ramaswamy, Phys. Rev. Lett. {\bf 63}, 1659 (1989).

\bibitem{cates} O. J. O'Loan, M. R. Evans and M. E. Cates, Phys. Rev. E.
{\bf 58}, 1404 (1998).

\bibitem{evans} M. R. Evans, J. Phys. A. {\bf 30}, 5669 (1997).

\bibitem{AM} A. Maritan, A. Rinaldo, R. Rigon, A. Giacometti and I. R.
Iturbe, Phys. Rev. E {\bf 53}, 1510 (1996); M. Cieplak, A. Giacometti, A.
Maritan, A. Rinaldo, I. R. Iturbe and J. R. Banavar, J. Stat. Phys. {\bf
91}, 1 (1998).

\bibitem{SC} S. N. Coppersmith, C.-h. Liu, S. N. Majumdar, O. Narayan and
T. A. Witten, Phys. Rev. E {\bf 53}, 4673 (1996).

\bibitem{krug} J. Krug and J. Garcia, cond-mat/9909034.

\bibitem{Melzak} Z. A. Melzak, {\it Mathematical Ideas, Modeling and
Applications, Vol II of Companion to Concrete Mathematics} (Wiley, New
York, 1976), p.271.

\bibitem{FF} P. A. Ferrari and L. R. G. Fontes, El. J. Prob. {\bf 3},
Paper no. 6 (1998).

\bibitem{IKR} S. Ispolatov, P. L. Krapivsky and S. Redner, Eur. Phys. J.
{\bf B2}, 267 (1998).

\bibitem{aldous} D. Aldous and P. Diaconis, Probablity Theory and Related
Fields, {\bf 103}, 199 (1995).

\bibitem{Haff} P. K. Haff, J. Fluid Mech. {\bf 134}, 401 (1983); S.
McNamara and W. R. Young, Phys. Fluids A {\bf 4}, 496 (1992); B. Bernu and
R. Mazighi, J. Phys. A {\bf 23}, 5745 (1990); E. Ben-Naim and P.
Krapivsky, cond-mat/9909176.

\bibitem{RM} R. Rajesh and S. N. Majumdar (unpublished).

\bibitem{GS} G. M. Schutz (preprint) (1999).

\end{thebibliography}
\end{document}